\documentclass[aps,prfluids,11pt,superscriptaddress,showpacs,floatfix,amsmath,amssymb,longbibliography]{revtex4-1}
\usepackage{bm}
\usepackage[overload]{empheq}
\usepackage{color}
\usepackage{url}
\usepackage{graphicx}
\usepackage{tabularx}

\begin{document}

\title{Scalar absorption by particles advected in a turbulent flow}

\author{A. Sozza}
\thanks{Corresponding author}
\email{asozza.ph@gmail.com}
\affiliation{Istituto dei Sistemi Complessi, CNR, via dei Taurini 19, 00185 Rome, Italy and INFN, sez. Roma2 ``Tor Vergata''}

\author{M. Cencini}
\thanks{Corresponding author}
\email{massimo.cencini@cnr.it}
\affiliation{Istituto dei Sistemi Complessi, CNR, via dei Taurini 19, 00185 Rome, Italy and INFN, sez. Roma2 ``Tor Vergata''}

\author{F. De Lillo}
\affiliation{Dipartimento di Fisica and INFN, Universit\'a di Torino, via P. Giuria 1, 10125 Torino, Italy.}

\author{G. Boffetta}
\affiliation{Dipartimento di Fisica and INFN, Universit\'a di Torino, via P. Giuria 1, 10125 Torino, Italy.}

\begin{abstract}
We investigate the effects of turbulent fluctuations on the Lagrangian statistics 
of absorption of a scalar field by tracer particles, as a model for nutrient uptake 
by suspended non-motile microorganisms. By means of extensive direct 
numerical simulations of an Eulerian-Lagrangian model we quantify, in terms of 
the Sherwood number, the increase of the scalar uptake induced by turbulence 
and its dependence on the Peclet and Reynolds numbers. Numerical results are 
compared with classical predictions for a stationary shear flow extended here to 
take into account the presence of a restoring scalar flux. We find that mean field 
predictions agree with numerical simulations at low Peclet numbers but are unable 
to describe the large fluctuations of local scalar uptake observed for large Peclet 
numbers. We also study the role of velocity fluctuations in the local uptake by 
looking at the temporal correlation between local shear and uptake rate and we find 
that the latter follows fluid velocity fluctuations with a delay given by Kolmogorov time scale. 
The relevance of our results for aquatic microorganisms is also discussed.
\end{abstract}

\date{\today}

\maketitle

\section{Introduction}
Mass or heat transfer in multi-phase systems is a problem of great
interest for both theory and applications. Several industrial
processes involve fluids with suspended particles that undergo
chemical reactions, where particles exchange mass or heat with the
surrounding fluid
\cite{harriott1962review,harriott1962mass,armenante1989}. Fluid
flows also mediate the uptake of nutrients and other biochemicals by
suspended (unicellular) microorganisms \cite{leal2007,neufeld2009}.
Previous works have shown that small-scale turbulence enhances the
transport of nutrients into the cell
\cite{karpboss1996,leal2007}. Such increase in the nutrient
flux is typically negligible for bacteria while it can be significant
for larger cells such as eukaryotic phytoplankton. However, these
studies have considered turbulence as an average, time independent,
shear flow while velocity and scalar fields in turbulent flows exhibit
strong fluctuations and develop large gradients that are completely
overlooked by a mean field description \cite{frisch1995}.
Consequently, the effects of turbulent fluctuations and their
intermittency on cellular uptake are largely unknown. By stirring the
nutrient patches, turbulence creates inhomogeneities and complex
landscapes of nutrient \cite{falkovich2001} that make the uptake
problem non-trivial. Moreover, the variability induced by
small-scale turbulence may affect the ecological strategies in terms
of growth and reproduction rates \cite{margalef1978}.

Previous numerical simulations have studied the problem of cellular
uptake using different approaches. For instance, continuous (Eulerian)
models for describing the concentration of both nutrients and (a
population of) absorbers have been used for studying phytoplankton
dynamics \cite{abraham1998,lopez2001} and for
quantifying the interaction between motility and turbulence in
bacterial chemotaxis \cite{taylor2012,desai2018}, but
do not provide information on the uptake by a single particle. 
Discrete, Lagrangian models have also been used for computing 
the uptake of nutrient in a still fluid or in simple laminar flows 
\cite{musielak2009,lambert2013}, thus 
disregarding the unsteadiness of turbulent flows.

In this work, we numerically study the effects of turbulent
fluctuations on nutrient uptake by using a mixed Lagrangian-Eulerian
approach. The nutrient is represented by a continuous, passive scalar
field, while the absorbers are represented by Lagrangian
particles. Both the nutrient and the absorbers are transported by a
realistic turbulent flow obtained by the integration of the
Navier-Stokes equations at high resolution. Scalar absorption is
implemented by volumetric sinks centered at the particle positions,
which has been shown to give accurate results in the absence of a flow
or in simple laminar flows \cite{sozza2018}. In order to
maintain a statistically stationary state in a finite volume, nutrient
is replenished by a uniform source (a chemostat, which models the
upwelling from a nutrient rich reservoir \cite{abraham1998,pasquero2005}). 
For an accurate comparison with simulations, we extend the analytical 
results of the uptake enhancement by a shear flow \cite{batchelor1979} 
to the presence of a restoring flux.

The remaining of the manuscript is organized as follows. In
Sec.~\ref{sec2}, we introduce the model and briefly describe the
analytical derivation of nutrient uptake in the presence of a
uniform restoring source. In Sec.~\ref{sec3}, we summarize
the numerical implementation of the model and the parameters
used in the simulations. In Sec.~\ref{sec4}, we discuss the numerical
results and their comparison with the analytical prediction based on
the mean field model. Finally, Section~\ref{sec5} is devoted to
conclusions and discussions. The appendices detail the analytical results.

\section{Mathematical models and theoretical results}
\label{sec2}
\subsection{Model equations}
We consider the general problem of $N$ discrete particles -- the
absorbers -- transported by an incompressible velocity field
$\bm{u}(\bm{x},t)$ together with a passive scalar field $c(\bm{x},t)$
-- the nutrient concentration. Particles are considered
as point tracers whose position $\bm{X}_i$ evolves according to
\begin{equation}
\dot{\bm{X}_i} = \bm{u}(\bm{X}_i,t)\,, \qquad i=1,\ldots,N\,.
\label{eq:2.1}
\end{equation}
The velocity field is a solution of the incompressible Navier-Stokes equation
\begin{equation}
\partial_t \bm{u}+ \bm{u} \cdot \bm{\nabla} \bm{u} =
- \bm{\nabla}p + \nu \triangle \bm{u} + \bm{f},
\label{eq:2.2}
\end{equation}
where $\nu$ is the kinematic viscosity, $p$ the pressure and $\bm{f}$
a body forcing injecting energy at rate $\varepsilon$ on average equal
to the energy dissipation rate, so to establish a statistically steady
turbulent state. The nutrient is advected by the flow, diffuses with
diffusivity $D$, is absorbed by the particles and uniformly restored
at a rate $\mu$ to maintain, on average, a constant concentration
$c_0$; its evolution reads
\begin{equation}
\partial_t c 
+ \bm{u} \cdot \bm{\nabla} c = D \triangle c
-\sum_{i=1}^{N} \beta_i \delta(\bm{x}-\bm{X}_i)\,c - \mu (c-c_0)\,.
\label{eq:2.3}
\end{equation}
In applications to plankton into the ocean, the uniform nutrient flux 
models the vertical advection from a deep reservoir 
-- a chemostat -- with constant concentration \cite{pasquero2005}. 
Following Ref.~\cite{sozza2018}, absorption from the $i^{th}$ particle 
is modeled through a volumetric absorption rate, $\beta_i$, and not 
via absorbing boundary conditions at the particle surface. 
According to this model, the instantaneous uptake of $i^{th}$ particle is 
given by 
\begin{equation}
\kappa_i(t) = \int d^3\bm{x} \beta_i \delta(\bm{x}-\bm{X}_i) c(\bm{x},t)\,.
\label{eq:2.4}
\end{equation}
The numerical implementation of this model has been calibrated and
tested by using configurations of one, two or more absorbers in still
fluid and a laminar shear flow for which analytical results are
available \cite{sozza2018}. Our main interest here is to quantify
the effect of turbulence on the uptake rate at the level of the single particle. 
Specifically, by denoting with $\kappa_\mu$ the asymptotic uptake rate 
obtained with the same chemostat but in the absence of the flow 
(i.e. when only diffusion is at play), we aim at quantifying the statistics 
of instantaneous Sherwood number, defined as 
${\rm Sh}(t)=\kappa_i(t)/\kappa_\mu$ and its average, 
and how they depend on the relevant parameters of the problem.

In the absence of a flow (see App.~\ref{appA}), the effect of the
chemostat is to exponentially cut-off the modification of the
concentration field with a \textit{screening length}
$\xi=\sqrt{D/\mu}$, therefore reducing diffusive interactions between
particles that without the chemostat are long ranged. As a consequence,
the usual Smoluchovsky rate at $\mu=0$, $\kappa_s=4\pi D Rc_0$, for an
absorbing spherical particle of radius $R$ is modified into
Eq.~(\ref{eq:a3}), which we rewrite here
\begin{equation} 
\kappa_\mu = \kappa_s( 1 + {R}/{\xi})\,.
\label{eq:2.5} 
\end{equation} 
By inverting (\ref{eq:2.5}), we obtain that 
a particle absorbing the nutrient with rate $\kappa_\mu$ has radius
\begin{equation}
R = \dfrac{\xi}{2}\left( \sqrt{1+\dfrac{\kappa_\mu}{\pi D \xi c_0}} -1 \right)\,,
\label{eq:2.6}
\end{equation}
which can be used to define an effective radius for the point-particle model.

The instantaneous particle Peclet number, quantifying the importance
of advection by the flow over diffusive transport, is then defined as
${\rm Pe}(t)=\gamma R^2 /D$ where $\gamma(t)$ measures the
instantaneous turbulent shear rate at the particle position defined as
$\gamma = (2 S^2)^{1/2}$, where $S_{ij} = \frac{1}{2}(\partial_i u_j+
\partial_j u_i)$ is the symmetric velocity gradient tensor at the
particle position. It is useful to consider also the nominal Peclet
number ${\rm Pe}_{\eta}=\gamma_\eta R^2/D$ where
$\gamma_\eta=1/\tau_\eta$ is the inverse of the Kolmogorov time
$\tau_\eta=(\nu/\varepsilon)^{1/2}$. We remark that, since $\gamma$
is a concave function of the energy dissipation rate, due to Jensen
inequality we have $\langle \gamma \rangle \le \gamma_{\eta}$ and
therefore $\langle {\rm Pe} \rangle \le {\rm Pe}_{\eta}$. We also
notice that, by introducing the Schmidt number ${\rm Sc}=\nu/D$ and
the Kolmogorov length $\eta=(\nu^3/\varepsilon)^{1/4}$, the nominal
Peclet number can be expressed as ${\rm Pe}_\eta=(R/\eta)^2 {\rm Sc}$.
The latter expression shows that, since the model requires $R \le
\eta$, the maximum attainable value of ${\rm Pe}_\eta$ is given by ${\rm Sc}$.
In the following, with some abuse of notation, when there is no ambiguity, 
we will often indicate the average Peclet and Sherwood numbers 
as ${\rm Pe}$ and ${\rm Sh}$, respectively.

\subsection{Theory of nutrient uptake in the presence of a chemostat}
Classical results on the effect of a flow on nutrient uptake, obtained 
assuming a constant concentration at infinity, 
predict two different regimes for small and large ${\rm Pe}$ \cite{karpboss1996},
\begin{subequations}
\label{eq:sh}
\begin{empheq}[left={{\rm Sh}=\empheqlbrace\,}]{align}
& 1 + 0.28 \,{\rm Pe}^{1/2} \qquad {\rm Pe} \ll 1
\label{eq:sh1} \\
& 0.55\,{\rm Pe}^{1/3} \qquad\quad\;\;\, {\rm Pe} \gg 1
\label{eq:sh2}
\end{empheq}
\end{subequations}
obtained respectively assuming a linear shear flow with a point sink
\cite{batchelor1979,frankel1968} and making use of boundary-layer theory
\cite{batchelor1979,levich1962}. 
In the following, we briefly show how the small ${\rm Pe}$ result (\ref{eq:sh1}) 
can be generalized to the presence of a chemostat, details of the computation 
can be found in Appendix~\ref{appB}. Note that the computation is based on 
a constant shear flow so that ${\rm Pe}$ should be interpreted 
as the average Peclet number, which in this case coincides with the instantaneous one.

We consider the concentration field $c(\bm{r},t)$ relative to the
center of a particle of radius $R$. The boundary conditions are
$c(R,t)=0$ (perfect absorption on the particle surface) and 
$\lim_{r \to \infty} c(\bm{r},t)=c_0$. The main effect of the velocity field
$\bm{u}$ is to change the uptake rate by deforming the shape of the
concentration profile with respect to the purely diffusive case.
Following Ref.~\cite{batchelor1979}, we consider a
particle smaller than the smallest scale in the flow
(i.e. $R<\eta$), so that the velocity field around it can be
expressed as a linear shear. We decompose the concentration field
into a mean profile and a fluctuating one $c'(\bm{r},t)$ that represents 
the deviations from the diffusive, spherical symmetric solution 
\begin{equation}
c(\bm{r},t) = c_0 \left( 1 - \frac{\kappa(t)}{4\pi D r c_0 (1+R/\xi)} e^{-(r-R)/\xi} \right) + c'(\bm{r},t)
\label{eq:3.1}
\end{equation}
where $\kappa(t)$ represents the total (still unknown) flux to the particle.

In the absence of a flow ($\bm{u}=0$) we have $c'=0$ and $\kappa=\kappa_{\mu}$, 
as given by Eq.~(\ref{eq:2.5}). 
In the presence of a flow, the relative increase of nutrient uptake, 
given by the Sherwood number ${\rm Sh}(t)=\kappa(t)/\kappa_{\mu}$, is 
readily obtained imposing the condition $c(R,t)=0$ in (\ref{eq:3.1}) 
which gives ${\rm Sh}(t)=1+c'(R,t)/c_0$.
The asymptotic (and here averaged) value of the Sherwood number, 
in the limit $t \to \infty$ can be obtained by extending the analysis 
of \cite{batchelor1979}, which for ${\rm Pe} \ll 1$ yields (see App.~A for details)
\begin{equation}
{\rm Sh} = 1 + \chi(\alpha) {\rm Pe}^{1/2}
\label{eq:3.2}
\end{equation}
where
\begin{equation}
\chi(\alpha) =\dfrac{1}{(4\pi)^{1/2}} \int_0^{\infty} dz \, 
e^{-\alpha z} \dfrac{\sqrt{z}}{24+z^2}\,,
\label{eq:3.3}
\end{equation}
with $\alpha=\mu \tau_\eta$ the relative time scale between stirring and nutrient supply. 
For $\alpha=0$, Eq.~(\ref{eq:3.3}) recovers 
Batchelor's result, $\chi(0)=\sqrt{\pi}/(6^{1/4}4) \simeq 0.283$. 
For $\alpha>0$ the function decreases with $\alpha$, meaning 
that the effect of the chemostat is to reduce the contribution of stirring to the nutrient uptake. 
This is somehow expected since with a fast chemostat (with $\alpha=O(1)$) 
the nutrient around the particle is uniformly restored
before the flow deforms the iso-concentration surfaces.

\section{Direct numerical simulations\label{sec3}}

\begin{table}
\resizebox{0.7\textwidth}{!}{
\begin{tabular}{cccccccccccccccc}
\hline \hline 
Run & $M$ & 
$\mu$ & $\nu$ & $E$ & $U$ & $T$ & $\tau_\eta$ & $L$ & $\eta$ & 
$\ell_B$ & $\xi$ & ${\rm Re}_\lambda$ & ${\rm Sc}$ & $\alpha_\eta$ \\
\hline
A1 & $128$ & $0.2$ & $1.6 \times 10^{-2}$ & $0.59$ & $0.63$ & $5.9$ & $0.40$ & $3.70$ & 
$0.08$ & $0.025$ & $0.089$ & $38$ & $10$ & $0.08$ \\
A2 & $256$ & $0.2$ & $6.4 \times 10^{-3}$ & $0.65$ & $0.66$ & $6.5$ & $0.25$ & $4.28$ & 
$0.04$ & $0.013$ & $0.057$ & $66$ & $10$ & $ 0.05$ \\
A3 & $512$ & $0.2$ & $2.5 \times 10^{-3}$ & $0.67$ & $0.67$ & $6.7$ & $0.16$ & $4.48$ & 
$0.02$ & $0.006$ & $0.035$ & $109$ & $10$ & $0.03$ \\
A4 & $512$ & $0.3$ & $2.5 \times 10^{-3}$ & $0.67$ & $0.67$ & $6.7$ & $0.16$ & $4.48$ & 
$0.02$ & $0.006$ & $0.028$ & $109$ & $10$ & $0.05$ \\
\hline
B1 & $128$ & $0.2$ & $6.4 \times 10^{-3}$ & $0.64$ & $0.65$ & $6.4$ & $0.25$ & $4.18$ & 
$0.04$ & $0.04$ & $0.179$ & $65$ & $1.0$ & $0.05$ \\
B2 & $1024$ & $0.2$ & $3.9 \times 10^{-4}$ & $0.70$ & $0.68$ & $7.0$ & $0.06$ & $4.78$ & 
$0.005$ & $0.005$ & $0.045$ & $287$ & $1.0$ & $0.01$ \\
\hline \hline
\end{tabular}
}
\caption{Simulation parameters: 
Run index, resolution $M$, chemostat rate $\mu$, kinematic viscosity $\nu$, 
energy $E = \langle |\bm{u}^2| \rangle/2$, rms velocity $U = (2E/3)^{1/2}$, 
integral times cale $T=E/\varepsilon$, 
Kolmogorov time scale $\tau_\eta = (\nu/\varepsilon)^{1/2}$,
integral length scale $L=U\,T$, 
Kolmogorov length scale $\eta=(\nu^3/\varepsilon)^{1/4}$,
Batchelor length scale $\ell_B = \eta / {\rm Sc}^{1/2}$, 
screening length $\xi = (D/\mu)^{1/2}$, 
shear rate $\gamma_\eta = 1/\tau_\eta$, 
Taylor Reynolds number ${\rm Re}_\lambda= U^2(15/\nu \varepsilon)^{1/2}$, 
Schimdt number ${\rm Sc}=\nu/D$, and 
$\alpha_\eta = \mu \tau_\eta$. 
In all runs the energy injection rate is fixed at $\varepsilon=0.1$, and $k_f=1.5$.}
\label{tab:param}
\end{table}

We solve Eqs.~(\ref{eq:2.2}-\ref{eq:2.3}) via direct numerical
simulation (DNS) on a triply periodic cubic domain of side
$\mathcal{L}=2 \pi$ using up to $M^3 = 1024^3$ grid points with a
$2/3$ dealiased pseudo-spectral solver and $2^{nd}$ order Runge-Kutta
time marching. The forcing in Eq.~(\ref{eq:2.2}), acting only at
large scales (in the wave number shell $k\leq k_f$), is chosen in such
way as to maintain the energy input $\varepsilon$ constant. This is
obtained by taking ${\bm f}(\bm x,t) = \varepsilon \bm u(\bm
x,t)/2E_{k\le k_f} \Theta(k_f - k)$, where $\Theta$ is the Heaviside
step function and $E_{k\le k_f}$ the kinetic energy restricted to the
wavenumbers $\leq k_f$ \cite{lamorgese2005,machiels1997}. We ensure
that small-scale fluid motion is well resolved by imposing the
Batchelor length scale $\ell_B = \eta/\sqrt{{\rm Sc}}$ (the smallest
scale in the problem since ${\rm Sc} \ge 1$) to be at least of the
same order of the grid spacing, ($k_{max}\ell_B > 1.0$, where
$k_{max}=M/3$ is the maximum wave number available after dealiasing).
The velocity field is integrated until a statistically stationary
state is reached, then the concentration field is initialized to the
constant value $c_0$. We explored a range of Taylor-scale Reynolds
number (${\rm Re}_{\lambda} \approx 38-287$) with two choices of the
Schmidt number (${\rm Sc}=1,10$).
Table~\ref{tab:param} summarizes the main DNS parameters.

As for the solid phase, we seed $N$ particles uniformly in the domain 
and let them move according to Eq.~(\ref{eq:2.1}). 
The fluid velocity and its gradients (needed to estimate the shear rate) 
at particle positions are obtained via a $3^{d}$ order interpolation scheme. 
The $\delta$-function in Eq.~(\ref{eq:2.3}) is regularized by
a function $f({\bf x})$ with compact support, product of three
functions of a single variable 
\begin{equation}
f({\bm x})=\dfrac{1}{\Delta^3} \phi(x) \phi(y) \phi(z)
\label{eq:3.4}
\end{equation}
where $\Delta=\mathcal{L}/M$ is the grid size. The function $\phi$ is
chosen to be symmetric, positive, normalized and with compact support 
around its center. One convenient form, used in the present work,
is 
$\phi(x)=(1/n)[1+\cos(2 \pi x/(n \Delta)]$ for $|x| \le n \Delta/2$ 
with $n=4$ \cite{peskin2002}.
The uptake $\kappa_i$ of each particle is then computed from 
(\ref{eq:2.4}) with the $\delta$-function replaced by 
$f({\bf x})$. More details can be found in \cite{sozza2018}.

In order to explore different values of the Peclet number, we selected
different effective radii of the particles by tuning the absorption
rate $\beta$. In particular, the radius is calibrated by performing,
for each set of parameters, a diffusive simulation without flow and
with static particles. For each $\beta$, the asymptotic uptake rate
$\kappa_\mu$ is measured and Eq.~(\ref{eq:2.6}) is used to define the
particle radius \cite{sozza2018}.

To optimize the computational costs, several particles were integrated in each run. 
We remark that, in general, the presence of many particles in a finite domain
induces diffusive interactions which tend to reduce the single particle uptake rate 
\cite{galanti2016,bhalla2013,sozza2018}. Although this effect is relevant to and 
interesting for applications \cite{dorsaz2010,lavrentovich2013}, in this work 
we focus on the single particle absorption, and therefore on dilute concentrations 
such that diffusive interactions are negligible. In this respect, the chemostat, 
inducing a screening length $\xi$, reduces the interactions among particles.

We can exploit the knowledge of the screening length to estimate 
the number of particles to be used to minimize the diffusive interactions. 
The flow being incompressible, the particle distribution remains uniform. 
By assuming a random uniform distribution of $N$ particles in a cube domain of side 
$\mathcal{L}$, the probability density function (PDF) 
of nearest-neighbors distance, for small $r$ takes the form \cite{chandrasekhar1943}
\begin{equation}
P(r) = 4\pi r^2\rho\exp \left(-\frac{4}{3}\pi r^3 \rho \right)\,.
\label{eq:3.5}
\end{equation} 
The mean inter-particle distance is $\langle r \rangle = a/\rho_p^{1/3}$, 
with $a = \Gamma(1/3)/(36\pi)^{1/3}$, where $\rho_p=N/\mathcal{L}^3$ 
is the particle number density. By choosing, e.g., $\langle r \rangle = 8 \xi$ 
one has that the probability to find two particles at distance less than $2\xi$ is only $1\%$.

\section{Results\label{sec4}}

\begin{figure}
\centering
\includegraphics[width=0.9\textwidth]{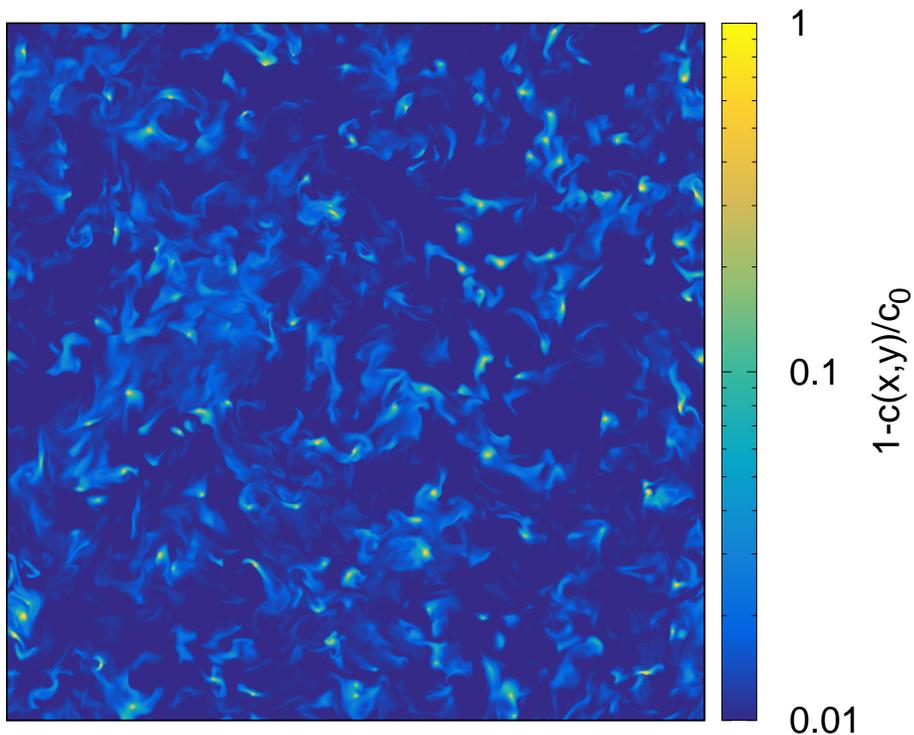}
\caption{(color online) Fluctuations of the concentration field
 $1-c(\bm{x},t)/c_0$ in a two-dimensional slab of $16$ grid
 points. Resolution $M=1024$ (Run B2).
\label{fig:sect}}
\end{figure}

We start by showing in Fig.~\ref{fig:sect} a typical example of the
concentration field in a two-dimensional section of the computational
box. Due to the absorption, small depletion zones are created around
the particles, which are then stretched by turbulence leading to
filament-like structures. The presence of these structures reflects
how turbulence locally increases scalar gradients, thus impacting
particle uptake.

\begin{figure}
\centering
\includegraphics[width=0.6\textwidth]{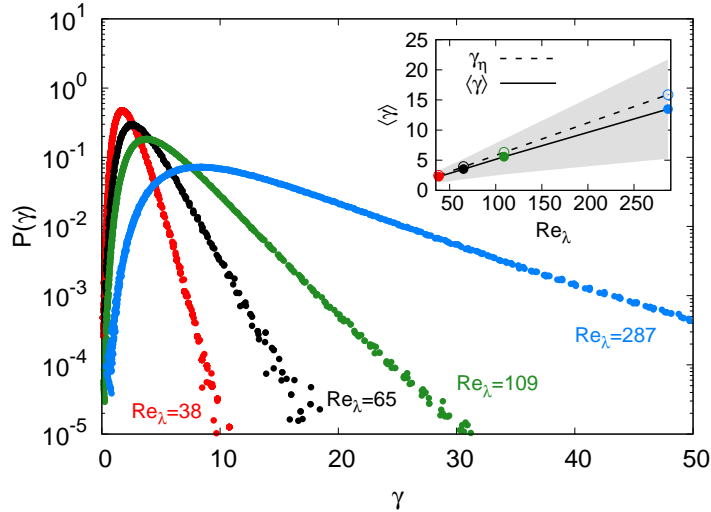}
\caption{(color online) Probability density function of shear rate 
$\gamma=(2 S^2)^{1/2}$ for different ${\rm Re}_\lambda$. 
Inset: mean shear rate $\langle \gamma\rangle$ (filled circles with solid line) 
and its root mean square (area in gray) compared with the dimensional 
estimation $\gamma_\eta = \tau_\eta^{-1}$ (empty circles with dashed line), 
due to the Jensen inequality $\langle\gamma\rangle < \gamma_\eta$.
\label{fig:pdf_gam}}
\end{figure}

In Fig.~\ref{fig:pdf_gam}, we plot the PDF of the shear rate $\gamma$
for different values of the Reynolds number ${\rm Re}_\lambda$. The
form of this distribution has been widely studied in previous works
and is characterized by non-Gaussian tails
\cite{biferale2008,buaria2019}, which become wider and wider at
increasing ${\rm Re}_\lambda$, the hallmark of intermittency in the
statistics of the velocity gradients.

\begin{figure}
\centering
\includegraphics[width=0.6\textwidth]{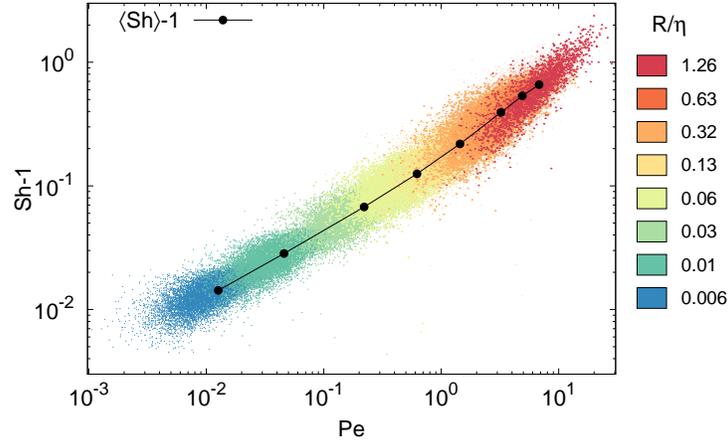}
\caption{(color online) Local gain of Sherwood number ${\rm Sh}-1$ 
versus local Peclet number ${\rm Pe}$ computed on the different particles of run A3, 
see Table~\ref{tab:param}. Colors refer to $9$ different values of the particles 
radius $R$ as in label. The solid line with filled circles represents the behavior 
of $\langle {\rm Sh} \rangle - 1$ versus $\langle {\rm Pe} \rangle$.
\label{fig:cloud}}
\end{figure}

Strong gradients are expected to cause local modification of the 
absorption. Indeed variations of $\gamma$ along the particle path 
modify the instantaneous value of the Peclet number. To understand and 
characterize these variations and their effect on the uptake we 
measure the instantaneous individual uptake $\kappa$, by using 
Eq.~(\ref{eq:2.4}), and shear rate $\gamma$ along each particle 
trajectories, in this way we can compute the local Peclet and Sherwood 
numbers. In Fig.~\ref{fig:cloud}, we plot the instantaneous value of 
${\rm Sh}-1$, i.e. the deviation from the diffusive uptake induced by turbulence, 
as a function of ${\rm Pe}$ for particles with $9$ different radii 
(each represented by a different color) transported by 
a turbulent flow at ${\rm Re}_\lambda=109$. Although a clear 
correlation between uptake rate and local shear is observed, as indeed 
the solid line shows that at changing the local Peclet number the Sherwood 
number changes according to the prediction valid for the average, we 
also observe large fluctuations of these values on a single particle (i.e. at fixed $R$).

\begin{figure}
\centering
\includegraphics[width=\textwidth]{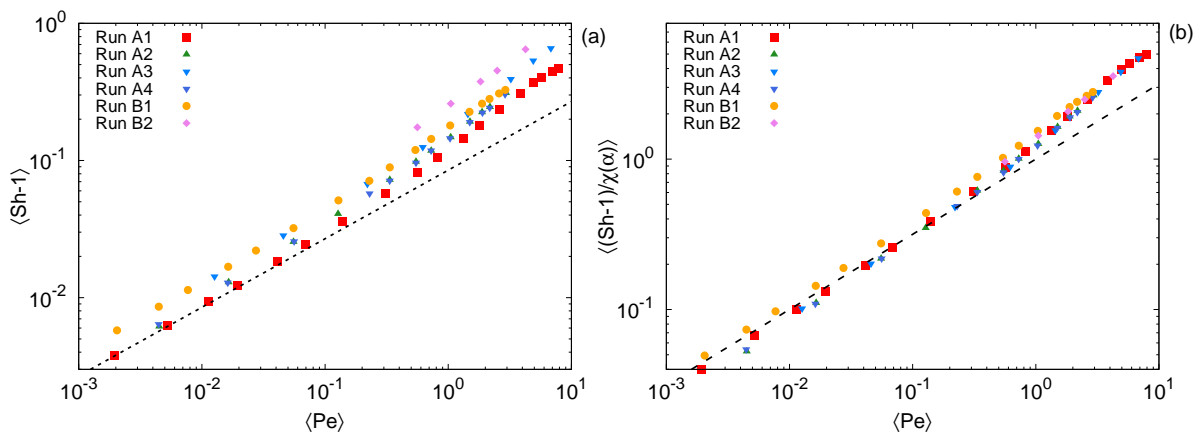}
\caption{(color online) Dependence of the mean Sherwood number on the 
mean Peclet number: (a) $\langle {\rm Sh}\rangle-1$ vs 
$\langle {\rm Pe}\rangle$ for all runs. Average is taken over all the particles 
with the same radius and over time, after discarding a transient. 
The dashed line displays the $\langle {\rm Pe}\rangle^{1/2}$ behavior 
of Eq.~(\ref{eq:sh1}). (b) $\langle {\rm Sh}\rangle-1$ rescaled by the coefficient 
$\chi(\alpha)$, see Eq.~(\ref{eq:3.3}). 
The dashed line represents the prediction (\ref{eq:3.2}).
\label{fig:ave}}
\end{figure}

The average values of the Peclet number $\langle {\rm Pe} \rangle$ 
and of the Sherwood number $\langle {\rm Sh} \rangle$ for the different 
simulations are shown in Fig.~\ref{fig:ave}a together with the 
classical theoretical prediction (\ref{eq:sh1}). Averages are 
computed over all particles having the same size and over time. 
Different symbols code the runs summarized in Table~\ref{tab:param}. 
The $\langle {\rm Pe}\rangle^{1/2}$ behavior of Eq.~(\ref{eq:sh1}) is 
clearly observable for small values of $\langle{\rm Pe}\rangle$. 
We do not observe the $\langle{\rm Pe}\rangle^{1/3}$ scaling of 
Eq.~(\ref{eq:sh2}), which is expected at larger $\langle{\rm Pe}\rangle$, 
however the points at largest $\langle {\rm Pe}\rangle$ of run A1 show 
a transition to a flatter scaling. As one can see, while the scaling 
Eq.~(\ref{eq:sh1}) is well reproduced, data obtained with different 
values of $\mu$ and ${\rm Re}_\lambda$ are not on the same master curve. 
The reason for this is the presence of the chemostat that 
modifies the constant in front of the $\langle{\rm Pe}\rangle^{1/2}$ scaling. 
In Figure~\ref{fig:ave}b we plot the Sherwood number rescaled with the 
coefficient $\chi(\alpha)$ (with $\alpha=\mu \tau_{\eta}$) given by 
(\ref{eq:3.3}), which generalizes Batchelor's result (corresponding to $\alpha=0$). 
As one can see, now we find a good collapse for all the runs characterized by 
different values of ${\rm Re}_\lambda$ and $\alpha$. 
For $\langle {\rm Pe}\rangle \lesssim 0.5$, the analytical prediction 
(\ref{eq:3.2}) provides an accurate description of the effect of 
turbulence on the uptake, which is mainly controlled by the Peclet number. 
We observe also some small difference between runs A and B, 
indicating a possible dependence on the Schmidt number 
which is not fully captured by the theoretical analysis. 
Remarkably, the collapse of the different curves is observed 
for all the available values of $\langle {\rm Pe}\rangle$, even beyond 
the range of validity of (\ref{eq:3.2}).

\begin{figure}
\centering
\includegraphics[width=\textwidth]{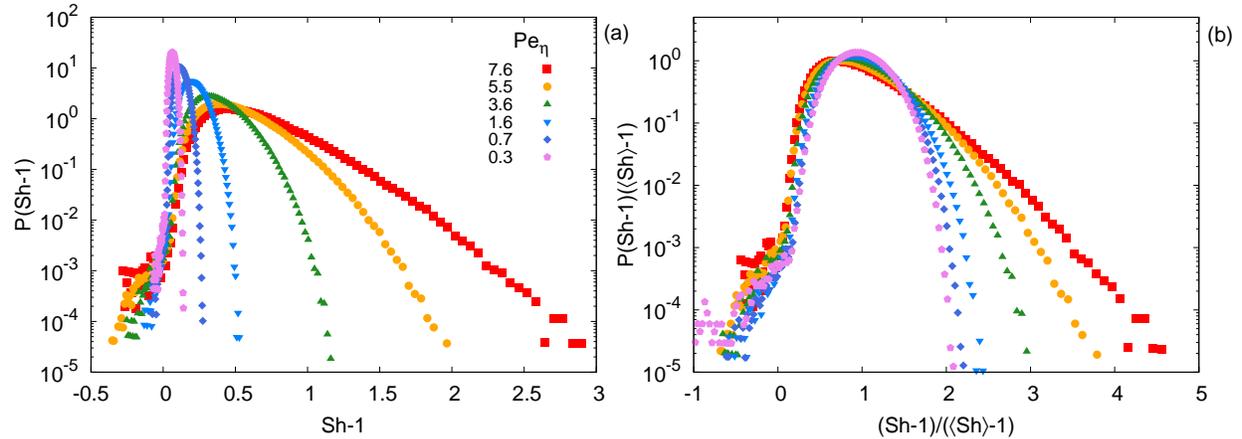}
\caption{(color online) PDF of the instantaneous Sherwood number 
${\rm Sh}$ at varying the nominal ${\rm Pe}_\eta$ number, 
i.e. for various particle radius: (a) Pdf of the deviation ${\rm Sh}-1$ 
of the Sherwood number from the diffusive value. 
Curves are plotted for different ${\rm Pe}_\eta$ and fixed 
${\rm Re}_\lambda=109$, ${\rm Sc}=10$ and $\alpha=0.05$ (run A3). 
(b) the same PDF of panel (a) normalized by their average. 
Note that the increased uptake by larger particles is 
also accompanied by more intense fluctuations.
\label{fig:pdf1}}
\end{figure}

Given the intense fluctuations that characterize turbulent gradients,
we expect the local uptake rate to be subjected to strong variations
with respect to the mean. In order to characterize those fluctuations
we study the PDF of the local Sherwood number ${\rm Sh}$ for different
values of the control parameters ${\rm Re}_\lambda$, ${\rm Pe}_\eta$,
${\rm Sc}$. In Fig.~\ref{fig:pdf1}, the distribution of ${\rm Sh}$ is
plotted, at fixed ${\rm Re}_\lambda$, ${\rm Sc}$ and $\alpha$, for
different values of ${\rm Pe}_\eta$ obtained by changing the effective
particle radius $R$. For very small values of ${\rm Pe}_\eta$ (hence,
small $R$), the observed values of ${\rm Sh}$ are confined to a narrow
interval around $1$. This confirms that the local uptake rate by small
particles is mildly influenced by turbulence both in terms of its
average and of its fluctuations. Increasing ${\rm Pe}_\eta$ (and
consequently, the particle radius), the distribution moves towards
larger values of ${\rm Sh}$ and develops wider right tails. This
change of shape in the distribution is made more evident in
Fig.~\ref{fig:pdf1}b, where the PDFs are normalized with the average
value $\langle {\rm Sh} \rangle$. 
The small left tails for ${\rm Sh}<1$ are due to the diffusive interactions 
among the particles in the simulation box. The relative importance 
of this effect is consistent with the estimation based on the particle number 
with random distribution and could, in principle, be eliminated by 
decreasing the number of particles. Nonetheless, the effect is very small and 
does not affect the global shape of the PDF.

\begin{figure}
\centering
\includegraphics[width=\textwidth]{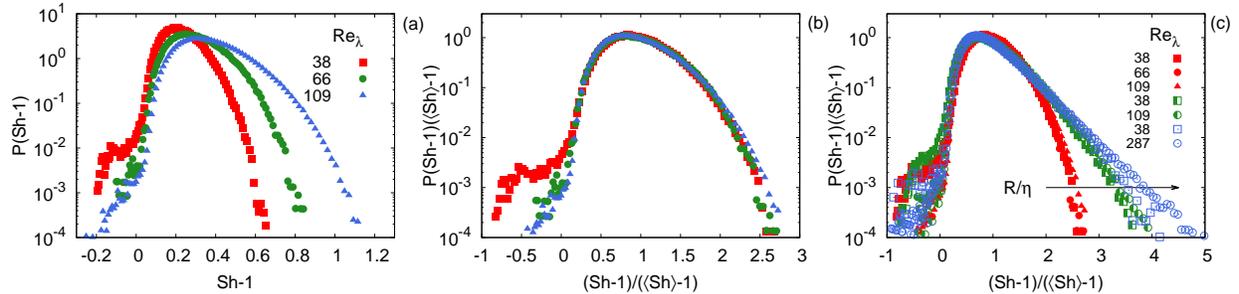}
\caption{(color online) PDF of ${\rm Sh}-1$ for different ${\rm Re}_\lambda$: 
(a) holding fixed ${\rm Pe}_\eta=3$, ${\rm Sc}=10$. The chemostat rate $\mu$ 
is also fixed so that larger ${\rm Re}_\lambda$ corresponds to smaller $\alpha$; 
(b) same as (a), but normalizing the PDFs of ${\rm Sh}-1$ with their average value, 
demonstrating that, even though the average $\langle {\rm Sh} \rangle$ increases 
with ${\rm Re}_\lambda$, the shape of the distribution is not strongly affected 
by the turbulence intensity. 
(c) same rescaling as in (b) repeated for three sets of particles, 
with different radii, namely $R/\eta=\sqrt{{\rm Pe}_\eta/{\rm Sc}}=0.57$ (red symbols), 
$0.87$(green symbols), $1.70$ (empty symbols) respectively. 
\label{fig:pdf2}}
\end{figure}

In order to directly scrutinize the effect of increasing turbulent
mixing, we now fix the particle size, i.e. the nominal Peclet number,
and consider different values of ${\rm Re}_\lambda$. In
Figure~\ref{fig:pdf2}a, we show the PDF of ${\rm Sh}$ for three cases
in which the nutrient replenishment rate $\mu$ of the chemostat is
kept constant while varying ${\rm Re}_\lambda$. By definition, for
fixed $\mu$, $\alpha$ decreases as turbulence becomes more intense and
this produces a shift of the PDF towards larger values of ${\rm Sh}$.
An increase in ${\rm Re}_\lambda$, however, does not seem to affect
the shape of the distribution, as one can appreciate from
Fig.~\ref{fig:pdf2}b, where the PDFs are normalized with their mean
values. In these three cases both the nominal Peclet, ${\rm Pe}_\eta$, 
and the Schmidt, ${\rm Sc}$, numbers are kept constant,
which physically speaking means that the ratio of the particle radius
to the Kolmogorov length is also constant, as $R/\eta=\sqrt{{\rm Pe}_\eta/{\rm Sc}}$. 
In Fig.~\ref{fig:pdf2}c, we show the PDFs for three different values of $R/\eta$. 
As one can see, the rescaled PDFs collapse fairly well, implying that $R/\eta$ 
dominates the overall shape of the distributions, especially the behavior 
of the right tail, with more intense fluctuations in uptake measured when 
the effective radius reaches the Kolmogorov scale. 
Residual effects in ${\rm Re}_\lambda$, however, cannot be completely ruled out. 
It is worth emphasizing that $R\sim\eta$ constitutes the upper limit 
for the particle size within our model, therefore the details of the 
statistics close to this limit should be taken with caution. 
However, the consistency of the behavior through about a factor 
three in radius seems to support the robustness of the observation.

\begin{figure}
\centering
\includegraphics[width=\textwidth]{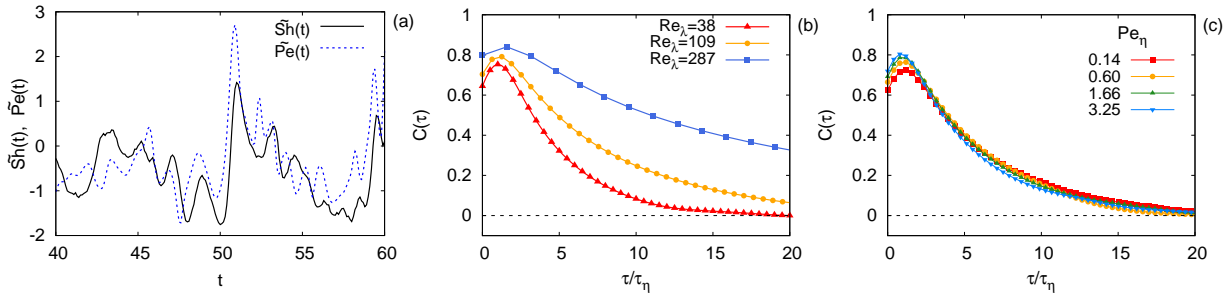}
\caption{(color online) Correlation between local strain and uptake rates: 
(a) Temporal signal of the normalized variables 
$\widetilde{{\rm Sh}}(t)=({\rm Sh}(t)-\langle {\rm Sh}\rangle)/\sigma_{\rm Sh}$ 
(black solid line) and 
$\widetilde{{\rm Pe}}(t)=({\rm Pe}(t)-\langle {\rm Pe}\rangle)/\sigma_{\rm Pe}$ 
(blue dashed line) for $Re_\lambda=66$ (Run A2) and $Pe_\eta=2$. 
(b) Temporal cross correlation $C(\tau)=\langle \widetilde{{\rm Pe}}(t) \widetilde{{\rm Sh}}(t+\tau) \rangle$ for different 
$Re_\lambda$ (Run A1,A3 and B2) while taking $\mu=0.2$ constant, 
thus changing $\alpha$, with $Pe_\eta=1$. Time is rescaled by $\tau_\eta$ 
to make clear that the maximum is attained at $\tau\approx \tau_\eta$. 
Without this rescaling the long time decay of the different curves is basically the same, 
meaning that it is mainly controlled by the chemostat rate $\mu$ (not shown). 
(c) for different ${\rm Pe}_\eta$, as in label, at fixed ${\rm Re}_\lambda=66$ 
and ${\rm Sc}=10$ (for clarity, not all data points are represented).
\label{fig:correl}}
\end{figure}

We conclude by briefly discussing the relevance of time correlations 
in the uptake process. Indeed, it is reasonable to expect that a local 
fluctuation of the velocity gradient at a given time should produce a 
corresponding fluctuation in the scalar uptake with some delay time. 
This is qualitatively confirmed by inspecting the temporal signals of 
the instantaneous Sherwood and Peclet numbers, shown in 
Fig.~\ref{fig:correl}a. In order to quantify this delay we compute the 
connected cross correlation between ${\rm Pe}$ and ${\rm Sh}$ 
normalized by their standard deviation, which is shown in 
Figs.~\ref{fig:correl}(b) and (c) at varying the relevant parameters. 
As one can see, the correlation functions attain their maximum value 
for a delay time of the order of $\tau_\eta$. 
The Kolmogorov times scale is indeed the time scale for 
the deformation of the nutrient field to take place around the 
particle in the viscous-diffusive regime of scalar transport, 
which is the one relevant to the problem \cite{brethouwer2003}.

\section{Conclusions\label{sec5}}

In this work we have studied the effect of turbulence on the scalar
uptake by spherical absorbing particles advected by the flow. By
means of realistic, direct numerical simulations at different
turbulent intensities, we computed the instantaneous absorption of a
scalar field, and its gain with respect to a purely diffusive process.
We used a point particle method with volumetric absorption of the
nutrient scalar field calibrated to represent particles of different
sizes (i.e. Peclet numbers).

The scalar uptake relative to a purely diffusive process, 
quantified by the Sherwood number ${\rm Sh}$, is found to depend on the 
average particle Peclet number ${\rm Pe}$, in agreement with classical 
predictions based on a mean-field representation of turbulence. 
Nonetheless, we observe strong fluctuations in the instantaneous 
and local value of ${\rm Sh}$, whose PDF develops wide tails in particular 
for large values of Peclet (i.e. large radii) and Reynolds numbers. 

By analyzing the time series of ${\rm Sh}$ and ${\rm Pe}$ along a
particle trajectory, we observed a delay between the two signals which
was quantified by computing the cross correlation function. This delay
is found to be of the order of the Kolmogorov time of the flow and
depends weakly on the Peclet number.

It is now worth discussing the possible relevance of our findings for
the nutrient uptake by small non-motile microorganisms transported by
fluid flows. In the ocean \cite{thorpe2007,williams2011}, using as
reference parameters $\varepsilon = 10^{-4}-10^{-8} m^2 s^{-3}$, 
$\nu = 10^{-6} m^2 s^{-1}$, and $D \sim 10^{-9} m^2 s^{-1}$ 
(for the most important nutrient like phosphate and nitrate) 
one has $\tau_\eta = 0.1 - 10 \, s$ and ${\rm Sc}=10^{3}$. 
For a phytoplankton cell of size $R = 10^{-6} m$ (e.g., a bacterium), 
the Peclet number ranges between ${\rm Pe} = 10^{-4} - 10^{-2}$ and, 
therefore, the effect of turbulence on nutrient uptake is expected to be negligible.
Conversely, for cells of size $R = 10^{-4} m$ we have ${\rm Pe} = 1 - 100$ 
and the average cellular uptake, according to our results, can be substantially 
affected by turbulence with an increase up to about two times with respect 
a purely diffusive environment \cite{karpboss1996}. 
In particular, our results show that, in this regime, the local value of the uptake 
can be much larger than the mean, with a PDF which develops very large tails.
To the best our knowledge, the effect of fast and strong nutrient fluctuations 
on phytoplankton growth is not known. 
Our findings suggest that it would be interesting to investigate this 
issue experimentally.

We also observe that the proposed model, 
besides being relevant to nutrient uptake of non-motile 
micro-organisms, is suitable to describe a variety of different 
applications, such as evaporation and condensation of droplets 
\cite{celani2005droplet,lanotte2009,siewert2017statistical}, 
in which the emission and absorption of supersaturated water vapor 
is mediated by turbulent transport.

The present analysis focused on the situation in 
which the nutrient is continuously replenished at all scales by a 
constant chemostat. While this can mimic nutrient upwelling 
from an underlying reservoir \cite{pasquero2005}, 
other kinds of forcing may be relevant to real applications. 
For instance, injections which, unlike the chemostat here used, 
replenish the nutrient only at large scales. 
Preliminary tests (not shown) in this direction \cite{Note1}
indicate that the main findings of our analysis are robust.
In particular, intense fluctuations in the local Sherwood number are 
observed independently on the forcing used, thus confirming the 
importance of a description that goes beyond the mean field approach.

Especially in the ocean, nutrient sources are often 
distributed in small ephemeral patches \cite{stocker2012}.
Thus, a natural extension of this model would be to consider a forcing 
for the nutrient not uniform in space and time, representing the variability 
present in nature. Other possible extensions of the model are in the direction 
of a better representation of the absorption mechanism which could take into
account more accurately of the local effects of the flow such as, for example, 
rotation of the cell due to local vorticity and its effect on the uptake \cite{batchelor1980}.

\section*{Acknowledgments}
We acknowledge HPC CINECA for computing resources (INFN-CINECA Grant No. INFN19-
fldturb). F.D. acknowledges PRACE for awarding access to GALILEO at CINECA through project
\emph{LiLiPlaTE}. G.B. and F.D. acknowledge support by the Departments of Excellence grant (MIUR).
A.S. acknowledges HPC CINECA for ISCRA-C Project \emph{NUPhTu} as well as Italian research project
MODSS (Monitoring Debris Space Stereo) Grant No. ID 85-2017-14966, funded by Lazio Innova
(Regione Lazio) according to Italian law L.R. 13/08.

\appendix

\section{Smoluchowski rate in a chemostat}
\label{appA}

Consider a spherical particle immersed in a quiescent nutrient concentration $c(\bm{x},t)$ 
sustained by a chemostat, i.e. ruled by the equation
\begin{equation}
\partial_t c = D\Delta c - \mu (c-c_0),
\label{eq:a1}
\end{equation}
with initial condition $c(\bm{x},0)=c_0$ and boundary 
conditions $c(R,t) = 0$ at the surface of the sphere and 
$c(\infty,t)=c_0$, where we used the spherical symmetry of the 
problem. At stationarity, the relative concentration, $\psi = 1 - c/c_0$, 
satisfies the equation $\psi'' +2\psi'/r - \mu \psi/D = 0$. 
Solving for the concentration yields
\begin{equation}
c(\bm{r},t) = c_0\left[1 - \dfrac{R}{r}e^{-(r-R)/\xi}\right]\,,
\label{eq:a2}
\end{equation}
where the long-range behavior of the solution without source term 
is exponentially damped with {\it screening length} $\xi=\sqrt{D/\mu}$. 
The uptake rate is obtained by integrating the nutrient flux 
$J = - D \partial_r c$ over the surface of the sphere
\begin{equation}
\kappa_\mu = \oint \bm{J}\cdot \bm{\hat{n}} ~ dS = \kappa_s\left( 1 + \dfrac{R}{\xi} \right)
\label{eq:a3}
\end{equation}
with $\kappa_s = 4\pi D R c_0$ being the usual Smoluchowski rate, 
which is recovered in the limit $\mu\to 0$ (i.e. $\xi\to \infty$).

\section{Generalization of the Batchelor calculations for a chemostat}
\label{appB}

We report here the details of the analytical derivation of the
Sherwood number behavior in the ${\rm Pe}\ll 1$ limit, by following
the work of Batchelor \cite{batchelor1979}, and extending his result
for the case of a chemostat.

We start by considering the equation for a nutrient concentration $c(\bm{x},t)$, 
advected by the velocity field $\bm{u}(\bm{x},t)$ and replenished by a chemostat with rate $\mu$,
\begin{equation}
\partial_t c 
+ \bm{u} \cdot \bm{\nabla} c = D \triangle c - \mu (c-c_0)\,.
\label{eq:B.0}
\end{equation}
Absorption by the particle is modeled by the boundary conditions: 
$c=0$ at the particle surface, i.e. for $r=R$ (where $r=|\bm{r}|$) 
and $c = c_0$ in the far away distance, $\bm{r}\rightarrow\infty$.

As a first approximation, we assume a time-independent linear shear
flow, i.e. $u_i = G_{ij} x_j$, with $G_{ij} = \partial_j u_i$
constant. As usual, the gradient tensor $G_{ij}$ can be written as
$G_{ij} = S_{ij} + \Omega_{ij}$, with the symmetric, $S_{ij}
=\frac{1}{2} (\partial_j u_i + \partial_i u_j)$, and anti-symmetric,
$\Omega_{ij} =\frac{1}{2} (\partial_j u_i - \partial_i u_j)$,
component representing straining motion and rigid-body rotation,
respectively.

We begin by searching for a solution in Fourier space for the
concentration field in the case of an instantaneous source with uptake
rate $\kappa$. In order to study the mass transfer in the proximity
of the particle, it is convenient to adopt the comoving coordinates
$\bm{r} = \bm{x}-\bm{X}_i$. We also rewrite Eq.~(\ref{eq:B.0}) for
the relative concentration $\psi = 1 - c/c_0$, with boundary condition
$\psi(\infty,t)=0$ and $\psi(R,t)=1$. Considering the Fourier
transform $\widehat{\psi}(\bm{q},t) = \int_{-\infty}^{\infty}
\psi(\bm{r},t) e^{-i\bm{q}\cdot\bm{r}} d^3\bm{r}$, Eq.~(\ref{eq:B.0})
reads
\begin{equation}
\dfrac{\partial \widehat{\psi}}{\partial t} - G_{ij} q_i \dfrac{\partial \widehat{\psi}}{\partial q_j} 
= - D q^2 \widehat{\psi} - \mu \widehat{\psi}
\label{eq:B.1}
\end{equation}
with $q=|\bm{q}|$.

The concentration field for a sustained source is obtained 
as the time integral of the solution of the instantaneous source, i.e.
\begin{equation}
\hat{\psi}(\bm{q},t) = \kappa \int_0^{t} ds ~ e^{ - D q_i B_{ij} q_{j} - \mu s}
\label{eq:B.2}
\end{equation}
with $B_{ij}(t)$ a time-dependent symmetric matrix that incorporates
the effects of the shear. In the absence of shear, it reduces to a
diagonal matrix that describes isotropic diffusion with a Gaussian
solution \cite{novikov1958,elrick1962}. Plugging the solution in 
Eq.~(\ref{eq:B.1}) yields the equation for the tensor $B_{ij}$ that, 
after some algebra, reads
\begin{equation}
\dfrac{d B_{ij}}{dt} = \delta_{ij} + G_{il} B_{jl} + G_{jl} B_{il}
\label{eq:B.4}
\end{equation}
Now anti-transforming, the concentration field becomes
\begin{equation}
\psi(\bm{r},t) = \kappa \int_0^{t} ds \int_{-\infty}^{\infty} \dfrac{d^3\bm{q}}{(2\pi)^3} \,
e^{i\bm{q}\cdot\bm{r} - D \bm{q} \cdot \bm{B} \bm{q} -\mu s}\,,
\label{eq:B.5}
\end{equation}
which can be easily solved by Gaussian integration, yielding in physical space,
\begin{equation}
\psi(\bm{r},t) = \dfrac{\kappa}{(4\pi D)^{3/2}} \int_0^t \!\!
\dfrac{ ds }{ \sqrt{ \det(\bm{B}) } } 
e^{ - \bm{r} \cdot \bm{B}^{-1} \bm{r}/(4D) -\mu s}\,.
\label{eq:B.6}
\end{equation}

The steady state solution, which corresponds to the diffusive
approximation around the absorbing particle, is obtained taking the
limit $t\rightarrow\infty$, and approximating the integrand with its
Taylor expansion in $r=0$ and $s=0$ \cite{frankel1968}. 
At the lowest order we can assume $B_{ij} = s \, \delta_{ij}$, 
$\det(B_{ij}) = s^3$ and $B_{ij}^{-1} = s^{-1} \delta_{ij} $, obtaining
\begin{equation}
\overline{\psi}(r) \approx \dfrac{\kappa}{(4\pi D)^{3/2}} 
\int_0^{\infty} \dfrac{ds}{s^{3/2}} e^{-\frac{r^2}{4Ds} - \mu s}
= \dfrac{\kappa e^{-r/\xi} }{4\pi D r}
\label{eq:B.7}
\end{equation}
The Sherwood number, defined as ${\rm Sh}(t) =1+c'(R,t)/c_0$ 
is computed subtracting the steady state solution 
to the global solution as $c'(\bm{r},t)/c_0 = \overline{\psi}(r) - \psi(\bm{r},t) $. 
We then perform the average over time by taking the limit $t\rightarrow\infty$, 
and then evaluate the integral at particle surface by taking the limit $r\rightarrow0$. 
Since $\kappa$ differs weakly from $\kappa_s$, we can approximate the integral as
\begin{equation}
{\rm Sh} = 1+ \dfrac{R}{(4\pi D)^{1/2}} \int_0^{\infty} 
\left( s^{-3/2} - \det(\bm{B})^{-1/2} \right) e^{-\mu s} ds
\label{eq:B.8}
\end{equation}

An approximation valid for a generic linear shear flows can be found
by expanding $B_{ij}$ as power series in $t$: $B_{ij} = \delta_{ij} t + B_{ij}^{(2)} t^2 + B_{ij}^{(3)} t^3 + \ldots$ .
Hence, substituting in Eq.~(\ref{eq:B.4}) we can determine the first coefficients
\begin{equation}
\begin{array}{l}
B_{ij}^{(2)}=\frac{1}{2}(G_{ij}+G_{ji})=S_{ij} \\[0.2cm]
B_{ij}^{(3)}=\frac{2}{3}S_{il}S_{jl} + \frac{1}{3} (S_{il}\Omega_{jl} + S_{jl}\Omega_{il} )
\end{array}
\label{eq:B.10}
\end{equation}
Now we consider the axes of reference to coincide with the principal 
axes of the rate of strain tensor $S_{ij}$ to obtain a simple shear 
flow along a preferential direction 
$G_{12} = \gamma$, $S_{12} = S_{21} = \gamma/2$, 
so that Eq.~(\ref{eq:B.10}) is satisfied by 
\begin{equation}
\begin{array}{c}
B_{11} = t(1+ \frac{1}{3} \gamma^2 t^2),\quad B_{22} = t,\quad B_{33}=t \\[0.2cm]
B_{12} = \frac{1}{2} \gamma t^2,\quad B_{21} = B_{23} = B_{33} = 0.
\end{array}
\end{equation}
The determinant of $B_{ij}$ is given by
\begin{equation}
\left( \dfrac{\det(\bm{B})}{t^3} \right)^{1/2} = 1+ \dfrac{\gamma^2 t^2}{24}
\end{equation}
By changing variable to $s=z/\gamma$ 
and by defining the parameter $\alpha = \mu/\gamma$, 
the Sherwood number in the case of a generic linear shear flow 
and in the presence of a chemostat is then given by
\begin{equation}
{\rm Sh} = 1 + \chi(\alpha) {\rm Pe}^{1/2}
\qquad \mathrm{with}
\qquad \chi(\alpha) =\dfrac{1}{(4\pi)^{1/2}} \int_0^{\infty} dz \, e^{-\alpha z} \dfrac{\sqrt{z}}{24+z^2} \,.
\end{equation} 
The integral admits a real solution for $\alpha>0$ 
and for $\alpha=0$ it recovers the prediction given by Batchelor, 
$\chi(0)=\sqrt{\pi}/(6^{1/4}4)$ \cite{batchelor1979}. 
The parameter $\alpha$ is the ratio between the time-scale 
of replenishment of the nutrient by the source 
and the stirring of fluid due to advection. 
Therefore, the nutrient source changes the rate of transfer 
and enters into the computation of the Sherwood number in a not trivial way.


\bibliography{biblio}

\end{document}